\documentstyle[aps]{revtex}
\begin{document}
\draft
\title{PLANE  -  CHAIN  COUPLING  IN  YBa$_{2}$Cu$_{3}$O$_{7}$ :
TEMPERATURE  DEPENDENCE  OF  THE  PENETRATION  DEPTH}
\author{X.\ Leyronas and R.\ Combescot}
\address{Laboratoire de Physique Statistique,
 Ecole Normale Sup\'erieure*,
24 rue Lhomond, 75231 Paris Cedex 05, France}
\date{Received \today}
\maketitle

\begin{abstract}
We have studied the penetration depth for a model of $YBa_{2}Cu_{3}O_{7} $  
involving pairing both in the $CuO_{2} $  planes and in the CuO  
chains. In this model pairing in the planes is due to an attractive  
interaction, while Coulomb repulsion induces in the chains an  
order parameter with opposite sign. Due to the anticrossing  
produced by hybridization between planes and chains, one obtains  
a d-wave like order parameter which changes sign on a single  
sheet of the Fermi surface and has nodes in the gap. We find  
that our model accounts quite well for the anisotropy of the  
penetration depth and for the absolute values. We reproduce  
fairly well the whole temperature dependence for both the  a  
 and the  b  directions, including the linear dependence at  
low temperature. We use a set of parameters which are all quite  
reasonable physically. Our results for the  c  direction are  
also satisfactory, although the situation is less clear both  
experimentally and theoretically.  \par 
\end{abstract}

\section {INTRODUCTION}
The debate about the mechanism of high $T_{c } $  superconductivity  
has seen recently marked progress by focusing on the symmetry  
of the superconducting order parameter. This question is intimately  
related to the mechanism but lends itself much more easily to  
experimental answers. A number of recent experiments, mostly  
on $YBa_{2}Cu_{3}O_{7} $  (YBCO), have shown that the order  
parameter displays lines of nodes on the Fermi surface. The  
most spectacular ones are the phase sensitive experiments \cite{1,2}  
which provide striking evidence for a change of sign of the  
order parameter at the Fermi surface. However these experiments  
are sensitive to the quality of the interfaces which are not  
under perfect control. It is therefore important that other  
experiments, including tunnelling \cite{3}, Raman scattering  
\cite{4} and penetration depth \cite{hardy93}, have shown the  
existence of low energy excited states. Although these experiments  
are not phase sensitive and could be explained by a strongly  
anisotropic s-wave order parameter, this interpretation seems  
fairly unlikely in view of the linear slope of the penetration  
depth at low temperature which is particularly striking and  
has been observed both in cristals \cite{hardy93} and in films  
\cite{la}.  \par 
\bigskip
A natural interpretation of the existence of nodes for the order  
parameter is the spin fluctuation mechanism \cite{6} which predicted  
a $d_{x^{2}-y^{2}} $ symmetry. But this hypothesis remains  
controversial because it meets with experimental as well as  
theoretical difficulties. The most prominent experimental stumbling  
block is the observation of a sizeable Josephson current \cite{7}  
in c-axis tunnelling between $YBa_{2}Cu_{3}O_{7} $ and  Pb which  
contradicts a pure d-wave symmetry, and which is hard to interpret  
as being caused by the small orthorhombic distorsion because  
of the large size of the effect. Other experiments display also  
a strong ab plane anisotropy for the superconducting properties,  
such as the magnetic field dependence of the specific heat anomaly  
at $T_{c } $  of the parent compound $LuBa_{2}Cu_{3}O_{7} $ . Most  
strikingly the penetration depth has a marked anisotropy \cite{basov}  
since, at zero temperature, the ratio between the superfluid  
densities $\lambda _{b }^{-2}(0)/\lambda _{a }^{-2}(0) $ is 2.3,
in good   agreement with the anisotropy of the normal state resistivity  
and of the square of the plasma frequency \cite{basov}. In $YBa_{2}Cu_{4}O_{8}
$, which has an orthorhombic distorsion of 0.8\% (that is somewhat  
less than $YBa_{2}Cu_{3}O_{7} $ which has 1.6\%) and where there  
are twice as many chains as in $YBa_{2}Cu_{3}O_{7}, $ the superfluid  
density anisotropy \cite{basov} is as high as 6.2 . In view  
of this evolution with increasing the number of chains, it is  
hard to escape the conclusion that the chains are conducting  
and that it is their direct contribution which is responsible  
for the superfluid density anisotropy. Now the linear term in  
$\lambda _{b }^{-2}(T) $ at low T is typically 3 times the  linear  
term in $\lambda _{a }^{-2}(T). $ If we attribute the linear term to  
the existence of nodes in the gap, this implies, roughly speaking,  
that there must be something like nodes on the chains. This  
does not fit easily in the spin fluctuation picture and is rather  
a strong indication that chains must be explicitely taken into  
account for a proper description of the superconducting properties,  
even if it is believed that they are unlikely to be the place  
for the dominant driving force toward superconductivity.   \par 
  \bigskip 
Another conspicuous problem of the spin fluctuation interpretation  
is the weak sensitivity of the critical temperature of $YBa_{2}Cu_{3}O_{7} $  
 to the presence of impurities. Indeed standard impurities produce  
in d-wave superconductors \cite{9} an effect analogous to pair-breaking  
by magnetic impurities in standard s-wave superconductors with  
a rapid decrease of the critical temperature with increasing  
impurity concentration following the Abrikosov - Gorkov law.  
However, except for Zn which is likely to have a magnetic environment  
once in an YBCO matrix, impurities seem to have a rather weak  
effect on the  $T_{c } $    of $YBa_{2}Cu_{3}O_{7} $  . Ion or  
electron irradiation have also shown a similarly weak sensitivity  
of the critical temperature.  \par 
  \bigskip 
In order to account for all these apparently contradictory experiments  
we have proposed recently a model \cite{10} where an order parameter  
with nodes is produced, not by in-plane repulsive interaction  
as in the spin fluctuation mechanism, but rather by a repulsive  
plane-chain pairing interaction which we attribute to mostly  
unscreened Coulomb repulsion. As a result the plane band and  
the chain band have order parameters with opposite sign. Since  
it is well known that these plane and chain bands hybridize  
because electrons are physically allowed to jump from planes  
to chains, this change of sign gives automatically rise to nodes  
of the gap in the region of the Brillouin zone where they anticross.  
In our model, although we do not exclude some pairing interaction  
coming from the chains, superconductivity arises primarily from  
an attractive interaction within the planes, due for example  
to phonon exchange. We can say that our order parameter is d-wave  
like because it is qualitatively similar \cite{10} to the 
$d_{x^{2}-y^{2}}$  model : on each sheet of the Fermi surface there  
are four nodes. On the other hand, since the chains play an  
essential role in our model, it is meaningless to consider an  
approximate tetragonal symmetry. Rather we have to consider  
only the orthorhombic symmetry, under which our order parameter  
is completely invariant (there is no symmetry breaking). Note  
that the $d_{x^{2}-y^{2}} $ order parameter is also invariant  
under this symmetry.  \par 
  \bigskip 
Since the order parameter of our model has nodes, it accounts  
naturally for all the experiments providing evidence for its  
change of sign and for the existence of low energy excitations.  
On the other hand, since our model is far from having the
$d_{x^{2}-y^{2}} $ symmetry because of the important role of the chains,  
it provides a simple explanation for the existence of the Josephson  
effect in c-axis  YBCO $   - $ Pb junctions and also naturally  
for the important anisotropy in the ab plane mentionned above.  
The existence of two weakly coupled bands  (plane and chain)  
in our model leads to the possibility of a weak sensitivity  
of the critical temperature to impurities, in agreement with  
experiment, as we have shown very recently \cite{11}. We note  
that our model has similarities with the two band S-N model,  
introduced by Abrikosov and Klemm \cite{abrkl} to account for  
Raman scattering data. It has also common features with single  
band models, where competing attractive and repulsive ( Coulomb  
) interactions lead to a change of sign of the order parameter  
within the band, such as the ones proposed by Abrikosov \cite{abr}  
and by Santi et al. \cite{santi}.  \par 
  \bigskip 
In this paper we will explore in detail the consequence of our  
model for the penetration depth. We note that, since it is a  
thermodynamic quantity, the penetration depth is less likely  
to be perturbed by extrinsic defects than a dynamical quantity.  
Moreover since one explores the sample over typically 1000 \AA,  
it is a bulk quantity. This is in contrast with surface experiments,  
such as photoemission or tunnelling, where only a few atomic  
layers are involved and which could give skewed informations  
if the surface happens to be somewhat different from the bulk  
for some reason. All these features make the study of the penetration  
depth particularly interesting. Since our model gives naturally  
nodes in the gap, it is clear that it gives rise qualitatively  
to a linear dependence of the penetration depth $\lambda (T) $ at low  
temperature. However it is not obvious that our model can account  
quantitatively for the size of the slope nor for the experimentally  
observed temperature dependence and its anisotropy. The purpose  
of the present paper is to explore these points in details. Here  
we will consider only  $YBa_{2}Cu_{3}O_{7} $  , since this compound  
is the best controlled and has been the most explored experimentally.  
Possible extension of our model to other compounds is left for  
future work.  \par 
  \bigskip 
Experiments on the penetration depth in YBCO, whether on cristals  
\cite{hardy93,mao} or on films \cite{la}, are now in reasonable  
agreement. In particular they find a linear behaviour at low  
temperature with a typical slope of 4.3 \AA/K for an average penetration  
depth $\lambda _{a b }. $
More recently the anisotropy of the penetration   
depth has also been measured in untwinned single cristals \cite{hardy94}.  
Since these last experiments provide stronger constraints on  
theory, we will mostly take them for comparison with experimental  
results ( note that the compound is more precisely $YBa_{2}Cu_{3}O_{6.95} $  
 but we will consider that it does not differ appreciably from  
$YBa_{2}Cu_{3}O_{7} $  ). The T = 0 values of the penetration  
depths are more difficult to obtain and we will make use of  
the results of Basov et al. \cite{basov} for comparison. However  
experimental agreement is not complete and there might still  
be some progress in the future on this side, with perhaps some  
surprises \cite{srikanth}. 

\section{THE  MODEL} 
We first introduce the essential ingredients of our model. The  
non interacting part of the Hamiltonian for the electrons in  
planes and chains is given \cite{10} by : 
\begin{eqnarray}
\label{eq1}
{\rm H}_{0}=\sum\nolimits\limits_{k}^{} 
{\varepsilon }_{k}{c}_{k}^{+}{c}_{k}+\sum\nolimits\limits_{k}^{} {\varepsilon
'}_{k}{d}_{k}^{+}{d}_{k}+\sum\nolimits\limits_{k}^{}
{t}_{k}{c}_{k}^{+}{d}_{k}+h.c.
\end{eqnarray}
where $c_{k }^{+} $ and $d_{k }^{+} $ create electrons in the plane  
and in the chain band respectively. The first term corresponds  
to an isolated plane while the second one describes isolated  
chains with respective dispersion relations $\epsilon _{k } $ 
and $\epsilon ^{\prime}_{k } $. The last term 
describes hopping between planes and chains.  
Actually since YBCO is made of stacks containing two $CuO_{2} $  
planes and one CuO-chains plane, this independent electron  
Hamiltonian does not correspond precisely to the situation found  
in YBCO. A more realistic description is obtained with the Hamiltonian:
\begin{eqnarray}
\label{eq2}
\matrix{{\rm H}_{0}=\sum\nolimits\limits_{n}^{}
 {\varepsilon }^{}{c}_{1,n}^{+}{c}_{1,n}+\sum\nolimits\limits_{n}^{}
{\varepsilon }^{}{c}_{2,n}^{+}{c}_{2,n}+\sum\nolimits\limits_{n}^{}
{t}_{p}({c}_{1,n}^{+}{c}_{2,n}+h.c.)\cr \cr +\sum\nolimits\limits_{n}^{}
{\varepsilon '}^{}{d}_{n}^{+}{d}_{n}+\sum\nolimits\limits_{n}^{}
{t}_{c}({c}_{1,n}^{+}{d}_{n}+h.c.)+\sum\nolimits\limits_{n}^{}
{t}_{c}({c}_{2,n}^{+}{d}_{n+1}+h.c.)\cr}
\end{eqnarray} 
Here all quantities are understood to depend on $k_{x } $  and  
$k_{y } $  and summations run also over $k_{x ,y }. $ Indices 1 and  
2 number the $CuO_{2} $  planes and the index  n  numbers the  
stacks. Introducing the even and odd plane band operators $c_{\pm } $  
by  $c_{1,2} $ = ( $c_{+} $ $\pm  $ $c_{-} $ $)/\surd 2 $ , 
and taking the Fourier   
transform in the  z  direction, we have:
\begin{eqnarray}
\label{eq3}
\matrix{{\rm H}_{0}=\sum\nolimits\limits_{k}^{}
 ({\varepsilon }^{}+{t}_{p}){c}_{+,k}^{+}{c}_{+,k}+\sum\nolimits\limits_{k}^{}
({\varepsilon }^{}-{t}_{p}){c}_{-,k}^{+}{c}_{-,k}+\cr \cr
+\sum\nolimits\limits_{k}^{} {\varepsilon '}^{}{d}_{k}^{+}{d}_{k}+
\sum\nolimits\limits_{k}^{}
{t}_{+}({c}_{+,k}^{+}{d}_{k}+h.c.)+\sum\nolimits\limits_{k}^{}
{t}_{-}({c}_{-,k}^{+}{d}_{k}+h.c.)(3)\cr}
\end{eqnarray}
Now a dependence and summation on  $k_{z } $  is understood, and  
we have set  t $_{+} $ = t $_{c } $ $\surd 2 $ cos ( $k_{z } $ c / 2 )
and t $_{-} $ = t $_{c } $ $\surd 2 $ sin ( $k_{z } $ c / 2 ) , with 
c  being    the size of the unit cell along the  z  direction. From band  
structure calculations \cite{yu,pickett,anders} the chain band  
Fermi surface crosses the odd plane band Fermi surface, whereas  
there is no crossing with the even plane band. Therefore the  
even plane band does not play an interesting physical role in  
our model. If it is omitted this leads back to the Hamiltonian  
Eq.(1) where the plane band is actually the odd plane band and  
$t_{k } $ = t $_{c } $ $\surd 2 $ sin ( $k_{z } $ c / 2 ) .
In order to simplify  
the discussion we will forget the even plane band. However when  
we will come to realistic calculations we will take into account  
its contribution.   \par 
  \bigskip 
With respect to pairing interactions, we take an attractive  
pairing in the plane while we assume a repulsive pairing between  
plane and chain. On the other hand we have no precise indication  
on the pairing interaction within the chains ( a simple and  
reasonable hypothesis would be to take the coupling constant  
equal to zero, but this is clearly somewhat arbitrary ). This  
leads to the effective interaction Hamiltonian ( with  g  and  
 K  positive ) :
\begin{eqnarray}
\label{eq4}
{\rm H}_{int}=-g\sum\nolimits\limits_{k,k'}^{} 
{c}_{k'}^{+}{c}_{-k'}^{+}{c}_{-k}{c}_{k}+K\sum\nolimits\limits_{k,k'}^{}
{d}_{k'}^{+}{d}_{-k'}^{+}{c}_{-k}{c}_{k}+h.c.-g'\sum\nolimits\limits_{k,k'}^{}
{d}_{k'}^{+}{d}_{-k'}^{+}{d}_{-k}{d}_{k}
\end{eqnarray}
 and the following mean field Hamiltonian :
\begin{eqnarray}
\label{eq5}
{\rm H}^{}={H}_{0}+\Delta \sum\nolimits\limits_{k}^{}
 {c}_{k}^{+}{c}_{-k}^{+}+\Delta '\sum\nolimits\limits_{k}^{}
{d}_{k}^{+}{d}_{-k}^{+}+h.c.
\end{eqnarray}
where the order parameter $(\Delta ,\Delta ^{\prime}) $ satisfies : 
\begin{eqnarray}
\label{eq6}
\rm \Delta =-g\sum\nolimits\limits_{k}^{}
 <{c}_{-k}{c}_{k}>+K\sum\nolimits\limits_{k}^{} <{d}_{-k}{d}_{k}>
\end{eqnarray}
\begin{eqnarray}
\label{eq7}
\rm \Delta '=K\sum\nolimits\limits_{k}^{} 
<{c}_{-k}{c}_{k}>-g'\sum\nolimits\limits_{k}^{} <{d}_{-k}{d}_{k}>
\end{eqnarray}
Because of hybridization, the plane and chain operators $c_{k } $  
and $d_{k } $ do not correspond to the eigenstates of $H_{0} $ and  
one has to perform a unitary transformation to diagonalize $H_{0} $  
 . The energies $e_{\pm }(${\bf  k}) are given by :
\begin{eqnarray}
\label{eq8}
\rm 2{e}_{\pm }(\bf k\rm )={\varepsilon }^{}(\bf k\rm )+
{\varepsilon '}^{}(\bf k\rm )\pm [({\varepsilon }^{}(\bf k\rm )-{\varepsilon
'}^{}(\bf k\rm ){)}^{2}+4{t}^{2}(\bf k\rm ){]}^{1/2}
\end{eqnarray}
When this transformation is carried out in the pairing part  
of the Hamiltonian one obtains interband pairing terms coupling  
the hybridized bands. However we will consider that  t $_{k } $  
is large enough so that these bands are well separated. Specifically  
this means that we assume $\Delta  $ , $\Delta ^{\prime} $  $ 
<< $ t $_{k } $ ( the parameters  
we use below do indeed satisfy this constraint, except for a  
small range around $k_{z } $  = 0 which is quantitatively unimportant  
). In this case one can easily see \cite{11} that these interband  
pairing terms are negligible. Therefore the transformation gives  
rise to a fully band-diagonalized Hamiltonian, with the order  
parameter in each band given explicitely by $\Delta _{\pm }(${\bf  k})  
= ( $\Delta  $ $\epsilon ^{\prime}(${\bf  k}) + $\Delta ^{\prime} $ 
$\epsilon (${\bf  
k}) )/( $\epsilon (${\bf  k}) + $\epsilon ^{\prime}(${\bf   
k})) ( {\bf k} being at the Fermi surface of the + or - band  
). The essential feature of our model is that, starting from  
isotropic interactions in planes and chains, we obtain a specific  
anisotropy for the order parameter ( $\Delta _{+}(${\bf  k}) ,
 $\Delta _{-}(${\bf    
k}) ). In particular we obtain an order parameter which changes  
sign and has nodes at the Fermi surface since we have managed  
to have $\Delta  $ and $\Delta ^{\prime} $ with opposite signs, 
while $\epsilon (${\bf  
k}) $/(\epsilon (${\bf   
k}) + $\epsilon ^{\prime}(${\bf  k})) goes essentially from 0 to 1 
when we move   
at the Fermi surface of a given band. Therefore we have an order  
parameter which is d-wave like, although we have assumed an  
attractive pairing in the planes.  \par 
  \bigskip 
 In order to obtain the temperature dependence of the penetration  
depth, we need the temperature dependence of the order parameter.  
This should be obtained as the solution of our two band model.  
As it is well known from the early work of Suhl et al. \cite{suhl},  
the resulting temperature dependence can be appreciably different  
from the standard BCS result. However we will not require in  
our calculations a ratio $|\Delta ^{\prime}| $ / $\Delta  $
between the chain and plane  
order parameters very different from unity ( it will be typically  
0.5 ). We have checked that, in the range of coupling constants  
which is of interest for us, within a weak coupling calculation,  
the departure from the BCS temperature dependence is rather  
small.  \par 
  \bigskip 
 Another complication is that, to obtain the order parameter  
and the penetration depth, we should perform a strong coupling  
calculation by solving Eliashberg equations. This is necessary  
to be consistent with the high value of the critical temperature  
( and indeed recent calculations of $T_{c } $  within the spin  
fluctuation model are also performed in the strong coupling  
regime \cite{monthoux}). This is also consistent with the rather  
high values of 2 $\Delta  $ / $T_{c } $  whenever it has been measured  
experimentally, by tunnelling or Raman scattering for example.  
And indeed we will make use below of values for 2 $\Delta  $ / $T_{c } $  
   somewhat larger than the BCS one. Such calculations do not  
make in principle any problem and they would be appropriate  
if we wanted precise quantitative results in order to fit experiment.  
However they have the disadvantage of introducing additional  
parameters - these would be the parameters necessary to describe  
the phonon spectrum \cite{rcgv}, in addition to the coupling  
constants ( including intraband Coulomb repulsion ). However  
our purpose in the present paper is not to perform refined calculations,  
but rather to show that our model can give a reasonable quantitative  
account for experiments without looking for perfect agreement  
( although as we will see we come already rather close to it  
). In addition we will find other sources of uncertainty. In  
this spirit we will take the following simplifying approach,  
as it is often done in the literature. We will perform a weak  
coupling calculation for the penetration depth and assume the  
weak coupling BCS result for the temperature dependence of the  
order parameter. We will take strong coupling effects into account  
only by a renormalization of the ratio 2 $\Delta  $ / $T_{c } $  . It is  
known that, for the moderate coupling constants we are interested  
in, strong coupling calculations do not give much different  
results. We will also invoke strong coupling effects when we  
will look for absolute values of the penetration depth since,  
because of mass renormalization, they will modify the bare results.  
The advantage of this approach is that it leaves us with only  
two adjustable parameters, namely the ratio 
$\Delta ^{\prime} $ / $\Delta  $ between   
the chain and plane order parameters ( we assume each one isotropic  
for simplicity as in our original model ) and the ratio 2 $\Delta  $  
/ $T_{c } $  . Moreover for both of these parameters we do not  
have physically much freedom. The price is a quantitative inaccuracy  
which we believe is rather small.

\section{IN-PLANE  PENETRATION  DEPTHS, LOW  TEMPERATURE  BEHAVIOUR}
Our calculations are made for clean superconductors. This limit  
seems appropriate for the experiments we consider, and more  
generally it seems to be the relevant one for high Tc superconductors  
because of their short coherence length. It would naturally  
be of interest to study the effect of impurities on the low  
temperature behaviour of the penetration depth, since this can  
help to discriminate between various theories. However since  
the physics is in this case controlled by the vicinity of the  
nodes of the gap, we do not expect much difference, at least  
qualiltatively, with what is found for the standard d-wave model  
\cite{hirsch}. On the other hand it is not clear that impurity  
effects can be entirely neglected in experiments, in particular  
for the penetration depth along the c-axis (we will take up  
this point again when we will look at this penetration depth  
). Nevertheless we do not consider them in the present paper.  \par 
  \bigskip 
As we have already mentionned, in order to account for the experimental  
results we will need a fairly sizeable value of the hybridization  
parameter  $t_{k } $  , which will be much larger than a typical  
order parameter. In this case the two hybridized bands are well  
separated and we have to consider a multiband situation. In  
such a case, since we can not use a free electron description,  
one may worry about the contribution of interband terms to the  
paramagnetic current as in Ref. \cite{atkcar}. However one can  
check that the superfluid density is zero at $T_{c } $ as it should  
( this results directly from the ''effective mass theorem'' \cite{asmer}  
), that is the paramagnetic current is exactly cancelled by  
the diamagnetic one. Below $T_{c } $ the modification of the interband  
terms due to the superconducting condensation is small provided  
the separation between the bands ( that is the difference between  
the energies of one-electron eigenstates with same wavevector  
) is large compared to the gap. More precisely, if D is a typical  
band separation and $\Delta  $ is a typical gap, the correction is of  
order ( $\Delta  $ / D ) $^{2} $ , and should be neglected for consistency  
in our case ( D is of order of the Fermi energy $E_{F } $ except  
in the anticrossing region where it is or order  $t_{k } $ , but  
since this region is small, the correction is of order $(\Delta  $ /  
$E_{F })^{2} $  anyway ).  \par 
  \bigskip 
 The expression of the penetration depth in this case is just  
the sum of the contributions from each band. This leads to the  
standard result along the i = x or y  axis:
\begin{eqnarray}
\label{eq9}
{{\rm \lambda }_{i}^{-2}(T) \over 2{e}^{2}{\mu }_{0}}=
\int_{\pm }^{}{{d}^{3}k \over {(2\pi )}^{3}}\delta ({e}_{\pm }){v}_{i,\pm
}^{2}2\pi T\sum\nolimits\limits_{0}^{\infty } {{\Delta }_{\pm }^{2}(T) \over
{[{\omega }_{n}^{2}+{\Delta }_{\pm }^{2}(T)]}^{3/2}}
\end{eqnarray} 
where the summation is over both the + and the - bands, and  
is limited to the Fermi surfaces by the Dirac function $\delta (e_{\pm }). $  
The gaps $\Delta _{\pm } $  , the energies $e_{\pm } $  and the velocities  
$v_{i ,\pm } $  are naturally all {\bf k} - dependent with the velocities  
$v_{i ,\pm }(${\bf  k}){\bf  =}
$\partial e_{\pm }({\bf    k})/\partial k_{i } $ 
 . The  
discrete summation in Eq.(9) is over the Matsubara frequencies  
$\omega _{n } $  = ( 2 n + 1 ) $\pi  $ T .  \par 
  \bigskip 
We will first consider the low temperature behaviour of the  
penetration depth. Indeed on one hand the most striking feature  
of the recent experimental results is the linearity of the low  
T dependence. On the other hand we can obtain in this regime  
explicit expressions in our model with a limited number of assumptions,  
because this dependence is controlled by the nodes of the gap  
which are located in the anticrossing region. We will obtain  
an analytical expression for this low temperature behaviour  
in the limit of small hybridization. With this aim we note that  
there is no linear term at low temperature if hybridization  
is zero. Hence if we subtract from Eq.(9) its counterpart with  
the hybridization parameter $t_{k } $  set to zero, the low T behaviour  
is unchanged. Then we are left with summations over quantities  
which are essentially nonzero only in the anticrossing region.  
If we assume  $t_{k } $  small enough, this region is small and  
we can perform a first order  $t_{k } $ expansion ( since the unhybridized   
term does not contribute to the low T, we can forget it in the  
rest of the calculation ). In this anticrossing region the Fermi  
velocities $\partial \epsilon ({\bf k})/\partial ${\bf  k} and 
$   \partial \epsilon ^{\prime}({\bf    k})/\partial ${\bf   
k} of the plane and chain bands can be taken as constants, equal  
to their value ${\bf v}_{p } $  and ${\bf v}_{c } $   at the crossing  
of the unhybridized Fermi surfaces.   \par 
  \bigskip 
It is more convenient to make a change of variables ( at fixed  
$k_{z } $  ) and take x = ( $\epsilon  $ - $\epsilon ^{\prime})/t_{k }
$  and y = ( $\epsilon  $ + $\epsilon ^{\prime})/2t_{k } $  
 as new variables instead of $k_{x } $  and $k_{y }. $ The Jacobian  
of the transformation is  
J = $\partial (x,y) $ / $\partial (k_{x } $ $,k_{y }) $ =  
$| $ ${\bf v}_{p } $  x ${\bf v}_{c } $  $| $ / $t_{k }^{2} $  .
The y integration  
is then easily performed. One can then see, by changing x  into  
 -x, that the contribution of the two bands + and - are identical,  
and we are thus left with the calculation on the - band.  The  
low T behaviour comes from the vicinity of the node $x_{0} $ of  
 $\Delta _{-} $ = (1/2) [ ( $\Delta  $ + $\Delta ^{\prime} $ ) $-
$ ( $\Delta  $ - $\Delta ^{\prime}  ) x / ( x^{2} $  
+4 $)^{1/2} $ ]. This is given by $x_{0} $  = ( $\Delta  $ - $| $
$\Delta ^{\prime}| $ )  
$/ $ ( $\Delta  $ $| $ $\Delta ^{\prime}| $ $)^{1/2}. $
Around this node, $\Delta _{-}   =  - 2  
( x - x_{0} $  ) ( $\Delta  $ $|\Delta ^{\prime}| $ $)^{3/2} $ /
( $\Delta  $ + $| $ $\Delta ^{\prime}| $ $)^{2}. $  
In performing the x integration, we obtain the dominant T dependence  
by evaluating $v_{i ,-} $  at the node $x_{0} $  . Since the gap  
$\Delta _{-} $  is now the only varying quantity, we are now led to  
a situation which is essentially identical to a d-wave calculation  
: after taking $\Delta _{-}/T $ as new variable, we are left with a  
purely numerical integral which turns out to be equal to  $\pi  $  
ln2 . Taking into account the two bands $\pm , $ we obtain for the  
low temperature dependence :
\begin{eqnarray}
\label{eq10}
\rm {\lambda }_{i}^{-2}(0)-{\lambda }_{i}^{-2}(T)=
{4 \ln 2 \over \pi }{{e}^{2}{\mu }_{0} \over {\hbar}^{2}}{\overline{t} \over
c}T{{(\Delta +\left|{\Delta '}\right|)}^{2} \over {(\Delta \left|{\Delta
'}\right|)}^{3/2}}{\left({{{v}_{i,-}^{2} \over J{t}_{k}^{2}}}\right)}_{0}
\end{eqnarray}
where  c  = 11.7 \AA\ is the size of the cell along the c axis,  
$\bar{\mathrm{t}}$ is the average of $|t_{k } $ $| $ over $k_{z } $  
and the value  
of  $(v^{2}_{i ,-} $  / J $t_{k } $ $^{2} $ $)_{0} $ at the node is  
given by :
\begin{eqnarray}
\label{eq11}
\rm {\left({{{v}_{i,-}^{2} \over J{t}_{k}^{2}}}\right)}_{0}=
{({\Delta }^{}{v}_{c,i}+{\left|{\Delta '}\right|}^{}{v}_{p,i}{)}^{2} \over
{(\Delta +\left|{\Delta '}\right|)}^{2}\left|{{\bf v}_{\rm p} \times 
{\bf v}_{\rm c}}\right|}
\end{eqnarray}
This dimensionless quantity is typically of order unity. Expressing  
numerically all known physical constants, this can be rewritten  
as :
\begin{eqnarray}
\label{eq12}
\rm {\lambda }_{i}(T)-{\lambda }_{i}(0)=
1.76{\lambda }_{i}^{3}(0){T \over {T}_{c}}{{T}_{c} \over \Delta
}{{(1+\left|{\Delta '}\right|/\Delta )}^{2} \over {(\left|{\Delta
'}\right|/\Delta )}^{3/2}}\overline{t}{\left({{{v}_{i,-}^{2} \over
J{t}_{k}^{2}}}\right)}_{0}
\end{eqnarray} 
where all the penetration lengths are now expressed in units  
of  1000 \AA\  and  the hybridization parameter  $\bar{\mathrm{t}}$
is expressed  
in eV. We want now to compare this result with experiments \cite{zhang}
on $YBa_{2}Cu_{3}O_{7} $ which gives a slope of 4.7 \AA/K in the  
along the  a  axis, and 3.6 \AA/K along the  b  axis.  \par 
  \bigskip 
A first point to make is that our result contains the zero temperature  
value $\lambda _{i }(0) $ of the penetration depth. We can naturally  
perform a calculation for this quantity, as we will do below.  
However the necessary ingredients to obtain numerical results,  
namely band structure informations and mostly renormalization  
due to the interactions, are not precisely known.
Moreover $\lambda _{i }(0) $   
is found by performing an averaging over the whole Fermi surface.  
Hence its physical content is essentially unrelated to the other  
ingredients of Eq.(12) for which only the vicinity of the nodes  
is relevant : they give a very local information while $\lambda _{i }(0) $  
contains a global one. Therefore we might just take $\lambda _{i }(0) $  
from experiment itself, which gives \cite{basov} the currently  
accepted values of $\lambda _{a }(0) $ = 1600 \AA\ 
and $\lambda _{b }(0) $ = 1000  
\AA  . This is all the more reasonable as one sees that our result  
is fairly sensitive to the value of $\lambda _{i }(0) $ since it comes  
in the third power.  \par 
  \bigskip 
Even taking $\lambda _{i }(0) $ from experiment we have many parameters  
in our result which are not so well controlled. We can first  
consider the anisotropy of the linear term, since most of the  
factors in Eq.(12) are common to the two axes. We have : 
 \begin{eqnarray}
\label{eq13}
\rm {{\lambda }_{y}(T)-{\lambda }_{y}(0) 
\over {\lambda }_{x}(T)-{\lambda }_{x}(0)}={{\lambda }_{y}^{3}(0)
 \over {\lambda}_{x}^{3}(0)}
{({\Delta}^{}{v}_{c,y}+{\left|{\Delta'}\right|}^{}{v}_{p,y}{)}^{2}
 \over ({\Delta}^{}{v}_{c,x}+{\left|{\Delta '}\right|}^{}{v}_{p,x}{)}^{2}}
\end{eqnarray} 
Since the chains are along the y direction, it is quite reasonable  
to assume that the component $v_{c ,x } $  of the chain electron  
velocity along x is essentially zero . Setting $v_{c ,y } $ =  
$v_{c } $  , we obtain the following numerical condition between  
our parameters in order to agree with experiment   :  $\Delta  $  $/ $  
 $\vert  $  $\Delta ^{\prime} $ $\vert  $    $+
 $  $v_{p ,y } $ / $v_{c } $ 
 = 1.77 $v_{p ,x } $ /  
$v_{c } $  . This relation is easily satisfied with gap ratios  
and velocity ratios of order unity, which is just what one would  
expect. It is therefore not a strong constraint. Just to give  
an example, the simple minded tight binding dispersion relations  
$\epsilon _{k } $ = $- $ 2 $t_{1} $ ( cos $k_{x } $  +
 cos $k_{y } $ )  $   - $ $\mu  $  
and  $\epsilon ^{\prime}_{k } $ =
 $- $ 2 $t'_{1} $ cos $k_{y } $ $- $ $\mu ^{\prime} $ 
for the plane  
and the chain bands give $\Delta  $  $/ $  $\vert  $
  $\Delta ^{\prime} $ $\vert  $ 
    = (1.77 sin  
$k_{x } $ / sin $k_{y } $ $- $ 1 ) $t_{1} $  / $t'_{1} $  . 
This is satisfied   
with  $t_{1} $  $\approx  $ $t'_{1} $  , a crossing located \cite{pickett}  
at $k_{x } $ $\approx  $ $\pi /2 $ and
 $k_{y } $ $\approx  $ $\pi /4 $ , and 
 $\vert  $  $\Delta ^{\prime} $ $\vert  $  $/ $  $\Delta  $  
 $\approx  $ 0.7 $. $  \par 
$ $  \par 
We consider now if we can account for the absolute value of  
the slope, say along the b axis. We assume again $v_{c ,x } $  
 = 0 and make use of the above relation between parameters imposed  
by the experimental anisotropy. The result can be considered  
as an equation giving the hybridization parameter  $\bar{\mathrm{t}}$. Taking  
$T_{c } $  = 93 K it reads (with  $\bar{\mathrm{t}}$ in eV):  \par 
\begin{eqnarray}
\label{eq14}
\overline{\rm t}=0.061{\Delta  
\over {T}_{c}}{\left({{\Delta  \over
\left|{\Delta'}\right|}}\right)}^{1/2}{{v}_{c} \over {v}_{p,x}}
\end{eqnarray} 
Taking as an example the values we have chosen in the preceding  
paragraph, together with the BCS isotropic weak coupling value  
$\Delta  $ / $T_{c } $ = 1.75, we obtain  $\bar{\mathrm{t}}$ = 0.09 eV.
However anisotropy   
as well as strong coupling are expected to raise $\Delta  $ / $T_{c } $  
, and if we rather take the average value found in experiments,  
namely $\Delta  $ / $T_{c } $  $\approx  $ 3 , this raises
$\bar{\mathrm{t}}$ up to 0.16 eV. However  
we stress again that the low temperature slope is sensitive  
to the physical parameters in the vicinity of the nodes. If  
we consider that the gap $\Delta  $  in the planes has actually some  
anisotropy, which is quite likely to happen, the gap value at  
the nodes can be lower than the maximum value that we have taken.  
Similarly we should take into account the local value of the  
mass renormalizations due to interactions, which would be quite  
hard to obtain. Hence it should be clear that we can not escape  
some appreciable uncertainty in our evaluations.   \par 
  \bigskip 
Nevertheless the typical value required for the hybridization  
parameter in order to account for the experiments, namely $\bar{\mathrm{t}}$
$\approx  $ 0.1 eV, is quite satisfactory, taking into account that we  
have basically no adjustable parameters. Indeed this parameter  
is small enough ( compared to the full band width of the plane  
or chain bands which is of order 1.6 eV ) to make our hybridization  
approach reasonable and consistent. In particular it is small  
enough for our model to explain \cite{11} the weak dependence  
of the critical temperature on impurity content, which is observed  
experimentally and which is one of the major problems of the  
spin fluctuation model. On the other hand one might worry that  
the hybridization we find is too large, although it is quite  
obvious from the start that we require a sizeable hybridization,  
since the linear dependence of the penetration depth at low  
temperature is a fairly strong effect. We will discuss later  
on this point together with our calculations of the penetration  
depth in the  c  direction and see that this not the case.

\section{CALCULATIONS  OF  THE  PENETRATION  DEPTH  OVER   
THE  WHOLE  TEMPERATURE  RANGE}
In the preceding paragraph we have discussed the low temperature  
behaviour of the penetration depth and we have found that our  
model can account for experiments in YBCO. The next interesting  
point is to know if it can do the same over the whole temperature  
range. A priori the answer is likely to be affirmative because  
the different components of this temperature dependence are  
controlled by different parameters of our model. Thus in principle  
they can be adjusted independently to reproduce the experimental  
results. Indeed the zero temperature value of the penetration  
depth is an average over the whole Fermi surface of band structure  
parameters (Fermi velocity, density of states), with proper  
renormalization due to interactions. Therefore it is a global  
quantity, which in particular does not depend on the values  
of the gaps.  On the other hand, as we have already seen explicitely  
above, the low temperature slope of the superfluid density depends  
only on the band structure parameters and the gaps in the vicinity  
of the nodes, which makes it a local quantity. Finally the slope  
of $\rho _{s} $ near $T_{c } $  is a global quantity
which depends both   
on band structure parameters and on the gaps on the whole Fermi  
surface. Since we consider the possibility of strong coupling  
effects, the temperature dependence of the gaps near $T_{c } $  
 is again an independent variable. The large number of parameters  
is an unpleasant feature of our model,  but it is likely to  
be also an inescapable ingredient of the physics of $YBa_{2}Cu_{3}O_{7} $  
.  \par 
  \bigskip 
Having realized that our model is quite flexible, we can still  
wonder if we can account for experiment with reasonable values  
of the parameters. In order to explore this problem conveniently  
we reduce flexibility by making some reasonable simplifying  
assumptions. We take the gaps constant within the plane band  
and the chain band. The ratio  R = $|\Delta ^{\prime}|/\Delta  $ 
between the gap $\Delta  $ in the plane band and the gap
$\Delta ^{\prime} $ in the chain band is one  
parameter of our calculation. We will assume that the temperature  
dependences of these two gaps are the same and take for it a  
simple renormalized BCS form : $\Delta (T) $ = ( $\Delta (0) $ / 1.76 $k_{B} $  
$T_{c } $ ) $\Delta _{BCS}(T). $
 The ratio $\Delta (0) $ / 1.76 $k_{B} $ $T_{c } $  
is another parameter of our calculation. For the band structure  
we take a simple tight binding ansatz which reproduces reasonably  
well the results obtained by band structure calculations 
\cite{pickett,anders},   
and in particular the Fermi surface. We take for the plane band  
the simple tight-binding dispersion relation $\epsilon _{k } $ = $- $ 2  
$t_{1} $ ( cos $k_{x } $  + cos $k_{y } $ ) +
2 $t_{2} $ cos $k_{x } $ cos   $k_{y } $    $- $ $\mu  $ ,
while we use naturally $\epsilon ^{\prime}_{k } $ 
= $- $ 2 $t'_{1} $  
cos $k_{y } $ $- $ $\mu ^{\prime} $ for the chain band.
Actually we find our
best    overall agreement for  $t_{2} $  $\approx  $ 0.
We note that the shape
of   the ( uncoupled ) odd plane band Fermi surface is indeed fairly  
well described by  $t_{2} $  $\approx  $ 0, 
except for the vicinity of $(\pi ,0) $   
and $(0,\pi ) $ which are missed in our simple description ( but  
since the velocities are small in these saddle point regions,  
this is likely to be unimportant for our purpose ). On the other  
hand for the even plane band a proper description of the Fermi  
surface requires a larger  $t_{2} $  .  \par 
  \bigskip 
Finally, in order to make realistic calculations, we have also  
included a contribution to the penetration depth from this even  
parity plane band, which is uncoupled to the chains. We note  
that including this contribution makes agreement with experiment  
more difficult to reach. Indeed, since we have no nodes on this  
piece of the Fermi surface, the contribution of this part has  
an  s-wave like temperature dependence and is essentially constant  
at low temperature. This makes it harder to account for the  
strong slope found experimentally. To perform the calculation,  
we need all the parameters characterizing this band as additional  
input. These parameters are not related in an obvious way to  
those of the odd parity plane band. Nevertheless we have made  
a simplifying assumption in order to avoid an increase in the  
number of parameters. The contribution of this plane band to  
$\lambda _{i }^{-2} $  seems isotropic (from band structure calculations  
\cite{pickett,anders}, its piece of the Fermi surface has essentially  
a tetragonal symmetry ). In the  a  direction there is no contribution  
from the chains in our model, and only the even and odd plane  
bands contribute. We have assumed for simplicity these contributions  
equal at zero temperature. This means specifically that we have  
taken $\lambda _{y }^{-2}(0)|_{even\ plane}
= \lambda _{x }^{-2}(0)| _{even\ plane}
= \lambda _{x }^{-2}(0)| _{odd\ plane
 +chain} $   . This fixes the T = 0 contribution   
of the even plane band. With respect to its temperature dependence,  
we take the gap of the even plane equal to the one of the odd  
plane band, namely $\Delta (T). $ These assumptions are quite reasonable,  
but it is clear that the even plane band is an additional source  
of flexibility of our model, if required.  \par 
  \bigskip 
Given the large number of parameters still left in our model,  
we have not proceeded to a systematic exploration in the parameter  
space. We have just made use of our results, obtained in the  
preceding paragraph, in order to locate the region in parameter  
space for a proper description of the low temperature behaviour,  
and we have then adjusted empirically the parameters to obtain  
a good averall agreement with experiment. It is possible that  
better sets of parameters can be found. However our purpose  
in this paper was to check if our model could give a satisfactory  
 account for experiment. Since our answer is positive we have  
not attempted an optimization of our parameters, which would  
be rather meaningless in view of the flexibility mentionned  
above. Later on we comment on the sensitivity of our results  
to the variation of our parameters. We find a good averall agreement,  
as shown in Fig.1, with the set of parameters  $t_{1} $ = 0.5  
eV, $t_{2} $ = $- $ 0.0 eV, $t'_{1} $ = 0.33 eV, $\mu  $ = $- $ 0.58 eV,  
$\mu ^{\prime} $ = $- $ 0.46 eV , $t_{c } $ = 0.2 eV, 
$|\Delta ^{\prime}|/ $ $\Delta  $ 
= 0.4 and 2 $\Delta  $  
/ $T_{c } $  = 2.2 .  \par 
  \bigskip 
Let us first consider the T = 0 values of the penetration depths.  
We find $\lambda _{a } $  = 800 \AA\  and $\lambda _{b } $ 
= 550 \AA.  A priori our  
model is not designed to calculate the absolute values of the  
penetration depths because it uses a rather primitive band structure.  
It is nevertheless interesting to compare our results to full  
band structure calculations \cite{mass} of the plasma frequencies  
$\Omega _{p i } $  which can be related to our penetration depths by  
 $\lambda _{i } $  = c / $\Omega _{p i } $   .
Their results give $\lambda _{a } $  =  
730 \AA, $\lambda _{b } $  = 450 \AA\  and $\lambda _{c } $ 
= 2100 \AA. We see that  
our values are rather near these ab initio calculations, which  
proves that our model is quite reasonable. In particular the  
anisotropy ratio they find is $\lambda _{b } $  / 
$\lambda _{a } $  = 1.62 while  
we obtain  $\lambda _{b } $  / $\lambda _{a } $  = 1.45.
Both are very near the  
experimental result 1.6 . This ratio does not vary much when  
we vary our parameters ( in a reasonable range). This is easy  
to understand. This ratio comes merely from the fact that a  
proper account has been made of the chain contribution to the  
penetration depth along the  b  axis. This agreement with experiment  
is a strong indication that the chains have to be included in  
order to get a proper description of the superconducting properties  
of YBCO as we have discussed in the introduction. Nevertheless  
both our calculation and Ref.\cite{mass} give absolute values  
for $\lambda _{i } $  which are much lower than the experimental results.  
The obvious explanation for this is mass renormalization, due  
to the interactions, which we have not taken into account. For  
an isotropic superconductor this would multiply the penetration  
depth by  ( 1 + $\lambda  $ $)^{1/2}, $ where $\lambda  $
is the coupling constant.   
In the present case this renormalization factor is most likely  
rather anisotropic, but since we have unfortunately no precise  
idea of this anisotropy, let us take it isotropic for the sake  
of this discussion . If we require for example that it pushes  
$\lambda _{a } $  to the experimental value $\lambda _{a } $
 = 1600\AA,  we find  
$\lambda _{b } $  = 1100 \AA, in quite good agreement with experiment.  
The corresponding value of $\lambda  $ is 3 which is reasonable for a  
strong coupling constant.  \par 
  \bigskip 
Let us turn now to our results for $\lambda ^{-2}(T) $ / $\lambda ^{-2}(0) $  
plotted in Fig.1 . As it can be seen we find a very reasonable  
agreement with the low temperature behaviour for both the  a  
 and the  b directions. However this requires the rather high  
value $t_{c } $ = 0.2 eV which we will discuss later on. The low  
temperature slopes are naturally sensitive to the other band  
structure parameters because the Fermi velocities in the anticrossing  
region must be high enough to account for experiment. This implies  
a proper location of the anticrossing region. However, once  
the proper range for these parameters is found, the results  
are not particularly sensitive to their choice in this range.  
They are however sensitive to the choice of $|\Delta ^{\prime}|/ $ 
$\Delta  $ because   
this parameter controls the position of the nodes, and the slopes  
are directly related to the orientation and the strength of  
the Fermi velocities at the nodes. We remark the value $|\Delta ^{\prime}|/ $  
$\Delta  $ = 0.5 that we use is quite reasonable physically : since  
we expect superconductivity to arise mostly because of interactions  
in the  planes, we should find the order parameter smaller in  
the chains than in the planes. We note also that the experimental  
results displays a linear behaviour over a remarkably large  
range of temperature since it goes almost up to T / $T_{c } $ =  
0.5 . A natural explanation for this feature is strong coupling  
effects which, by increasing 2 $\Delta  $ / $T_{c } $  , increase the scale  
for the low temperature behaviour (which is $\Delta  $ ) compared to  
the overall temperature scale (which is $T_{c } $  ). This effect is  
not fully accounted for by our calculations because we have  
a rather small 2 $\Delta  $ / $T_{c } $  = 2.2 . However a larger value  
would lower the low temperature slopes and give a disagreement  
with experiment. This situation is due to our simplifying choice  
of an isotropic gap. Choosing an anisotropic gap would make  
it possible for us to have a large slope ( related to the gap  
values near the nodes ) and a large linear region (related to  
the gap values farther from the nodes). Finally the agreement  
is not quite good when we get near  $T_{c } $  , but this was expected.  
Indeed experimental results \cite{kamal} give a ( 1 - T / $T_{c } $  
 $)^{y } $ dependence for $\lambda (T) $ near $T_{c } $
 with y $\approx  $ 1/3 ,  while  
our mean field approch leads us naturally to the standard  (  
1 - T / $T_{c } $  $)^{1/2} $  dependence. If this experimental  
result is confirmed and the effect is due to a large domain  
for critical fluctuations \cite{kamal}, including this phyics  
in our model will naturally bring us in better agreement with  
experiment.

\section{C-AXIS  PENETRATION  DEPTH} 
We come now to the calculation of the c-axis penetration depth.  
We note first that our simple model is intended to describe  
plane-chain coupling and is at the start a two-dimensional model.  
In order to make more realistic calculations, we have included  
a $k_{z } $ dependent coupling in its simplest form. However this  
description is obviously very crude and we may fear that it  
gives a poor account of transport in the  c  direction. For  
example our plane-chain coupling does not have any $(k_{x },k_{y }) $  
dependence. Physically we forget that plane-chain hopping is  
likely to happen through the orbitals of the O4 apical oxygen.  
Similarly we do not take due account of plane-plane hopping  
which is obviously necessary for transport in the  c  direction.  
Clearly a mismatch between plane-plane hopping and plane-chain  
hopping would strongly affect these transport properties ( it  
has for example been suggested by Xiang and Wheatley \cite{xiang}  
that the hopping along the  c  axis vanishes at the nodes ).  
It is nevertheless interesting to see what is the result of  
our model for the c-axis penetration depth, and we will see  
that it does not fare so badly.  \par 
  \bigskip 
Our raw value at T = 0 is $\lambda _{c } $  = 2250 \AA . 
This result compares   
quite well with the result $\lambda _{c } $  = 2100 \AA\ of band structure  
calculations given above. This is quite important because it  
validates the value of our plane-chain hopping parameter $t_{c } $  
 = 0.2 eV, which we might have feared to be too large. Therefore  
this value of $t_{c } $    is in reasonable agreement with what  
comes out of these band structure calculations. Next we have  
to renormalize this result as we have done above for the  a  
and b  directions. Taking the same renormalization factor, we  
obtain  $\lambda _{c } $  = 4500 \AA . This is clearly smaller than the  
experimental result  $\lambda _{c } $  $\approx  $ 10 000 \.
A although this is not  
way off the mark. We note that this same problem is shared by  
band structure calculations.   \par 
   \par   \bigskip 
It is perhaps not too difficult to understand this discrepancy.  
First the experimental side is probably less secure than for  
the in-plane measurements since $\lambda _{c } $ is not obtained directly,  
but requires a subtraction of the a or b axis contributions  
which might be difficult to make precisely. Moreover, although  
experiments agree \cite{mao,hardyc} to find $\lambda _{c } $ 
$\approx  $ 10 000\AA , they strongly disagree with respect to the temperature
dependence.   While Mao et al. \cite{mao} obtain at low temperature a linear  
decrease of $\lambda _{c }^{-2} $  much stronger than in the a or b  
direction, Bonn et al. find a much weaker dependence than for  
$\lambda _{a } $  or $\lambda _{b } $ .
On the theoretical side our calculation   
is performed in the clean limit. While this sounds quite reasonable  
for the in-plane directions, one may reasonably think that transport  
along the c axis is much more sensitive to any kind of scattering  
or disorder. The importance of incoherent hopping in the c direction  
has been stressed for example by Graf et al. \cite{graf} and  
Radtke et al.  \cite{radtke}. Clearly the weakness of coherent  
hopping in the c direction makes incoherent hopping much more  
relevant than for in-plane transport and this will lead to an  
increase in the penetration depth. Similarly since c axis transport  
implies electrons going through the chains, it will be much  
more sensitive to chain disorder than in-plane transport. This  
is confirmed by the strong dependence \cite{takenaka} of the  
c-axis resistivity on oxygen content in $YBa_{2}Cu_{3}O_{7-y } $  
 while the in-plane resisitivity depends much less on y. Naturally  
non stoechiometry brings easily a strong oxygen disorder in  
the chains, and because of the one dimensional nature of these  
chains, the electronic properties are much more perturbed by  
this disorder. A related explanation is the possible existence  
of layers of impurities or defects parallel to the planes, which  
would contribute to block c-axis transport. Such a situation  
would be similar to the one found in films for the ab penetration  
depth \cite{la} where extrinsic ''weak links'' were shown to increase  
the penetration depth from the intrinsic value of 1700 \AA\ up  
to 3500 \AA. The ratio between these two values is similar to  
the discrepancy between our result and the experimental one  
for $\lambda _{c }(0). $ At the same time, these weak links make the  
low temperature dependence weaker than the intrinsic one. It  
is tempting to attribute the difference between the low temperature  
behaviour found by Mao et al. and by Bonn et al. to such an  
effect, although this seems in contradiction with the fact that  
they find essentially the same value for $\lambda _{c }(0). $    \par 
  \bigskip 
Let us now consider the temperature dependence that we find  
for $\lambda _{c }^{-2}(T)/\lambda _{c }^{-2}(0). $
It can be seen in Fig.2   
that it is quite strong. The physical explanation is quite clear.  
In our model all the $k_{z } $ dependence of the band structure  
comes from hybridization, and hence it is stronger in the anticrossing  
region. This implies that the largest $v_{k,z } $  comes from  
this region. Since the nodes of the gap are in this same region,  
their influence are particularly strong and in particular the  
low temperature slope is large. We notice that our result is  
very analogous to the one found experimentally \cite{mao} by  
Mao et al. This is naturally satisfactory, but from the above  
discussion we should be careful not to draw any conclusion from  
this agreement.  \par 
  \bigskip 
We consider finally a slightly different version of our model  
which could correct for the discrepancy between our result and  
experiment for $\lambda _{c }(0). $ We may indeed wonder if the large  
value of $\lambda _{c }(0) $ is not due to a weak plane-plane coupling.  
This possibility has been suggested by Basov et al \cite{basov}.  
In order to explore our model in this direction, we assume that  
the plane-plane coupling  $t_{p } $   is small. To zeroth order  
one combination of the plane bands is uncoupled and has energy  
$\epsilon _{k } $  while the orthogonal combination of the plane bands  
and the chain band are hybridized in the usual way with their  
dispersion relation given by Eq.(8) with  $t_{k } $ =
$\surd 2 $ $t_{c } $. The in-plane penetration 
depths can be calculated to zeroth  
order. On the other hand the velocity along the  c  axis is  
zero at this order, and we have to go to first order in  $t_{p } $  
for the calculation of $\lambda _{c }. $
This gives $\hbar v_{k,z } $   
 /c =   $t_{p } \sin (k_{z }c) $  for the even plane band, and  
$\hbar v_{k,z } $  /c =   2 $t_{p } $   $t_{c }^{2} $ 
$   sin $ $(k_{z } $   
c)/ [ ( $e_{\pm } $ - $\epsilon  $ $)^{2} $ + 2  $t_{c }^{2} $  $    $ ] .
Therefore   
we can obtain the experimental value of $\lambda _{c }(0) $ by ajdusting  
$t_{p} $  . We show in Fig.3  the result of our calculations for  
the set of parameters  $t_{1} $ = 0.33 eV, $t_{2} $ = $- $ 0.0 eV,  
$t'_{1} $ = 0.4 eV, $\mu  $ = $- $ 0.7 eV, $\mu ^{\prime} $ = 
$- $ 0.56 eV , $t_{c } $   
= 0.2 eV, $|\Delta ^{\prime}|/ $ $\Delta  $ = 0.5 and 
2 $\Delta  $ / $T_{c } $ 
 = 3 . As it can  
be seen the agreement with experiment is quite good. For the  
 a  direction it is  slightly better than in Fig.1 because the  
average hybridization parameter is larger ( in the present case  
it is constant while in the preceding case it is proportional  
to  sin $(k_{z }c/2) $ ) and the ratio $\Delta  $ / $T_{c } $ 
is larger.   
We note that we find for $\lambda _{c }(T) $ a temperature dependence  
which is now in good agreement with Bonn et al. data \cite{hardyc}.  
Turning now to the absolute values, our raw results at T = 0  
are  $\lambda _{a } $  = 1300 \AA\  and $\lambda _{b } $  = 650 \AA.
If for 
example  
we renormalize $\lambda _{b } $    to 1000 \AA\ (and
$\lambda _{a } $  to 2000 \AA\  
), we find $\lambda _{c } $  = 10 000 \AA\ if we take  $t_{p } $  =
 0.043 eV. This small value justifies our first order expansion.

\section{CONCLUSION}
In this paper we have investigated the consequences of our coupled  
plain-chain model of $YBa_{2}Cu_{3}O_{7} $  for the penetration  
depth and compared them with experiments. We have found that  
our model accounts quite nicely for the anisotropy of the penetration  
depth in the  ab  plane at  T = 0 (and also for the absolute  
values, when mass renormalization is taken into account). This  
is obvious qualitatively since the chains contribute only in  
the  b  direction. But the agreement with experiment is also  
quite good quantitatively. We reproduce fairly well the linear  
dependence at low temperature and more generally the whole temperature  
dependence for both the  a  and the  b  directions (except near  
$T_{c } $  where experiments \cite{kamal} do not give a mean field  
dependence). All this is obtained with a set of parameters which  
are all quite reasonable physically. In particular we require  
naturally a sizeable hopping term between plane and chain to  
account for experiment, but the value we use is in agreement  
with results from band structure calculations. Regarding the  
penetration depth along the  c  axis, the situation is less  
satisfactory since two experiments strongly disagree with respect  
to the temperature dependence. We can actually reproduce both  
temperature dependence by making slightly different choices  
for our model to describe hopping in the  c  direction. The  
absolute value of the penetration depth disagrees typically  
by a factor 2 with experiments, but a natural explanation is  
the important effect of impurities and disorder on the transport  
properties along the  c  axis.

\section{ACKNOWLEDGEMENTS}
We are very grateful to W. Hardy for fruitful discussions on  
our model and to  N. Bontemps, P. Monod and L. A. de Vaulchier  
for discussions on the experimental situation.  \par
\bigskip
* Laboratoire associ\'e au Centre National
de la Recherche Scientifique et aux Universit\'es Paris 6 et Paris 7.

\begin{figure}
\caption{ a) $\lambda _{a }^{-2}(T) $ / $\lambda _{a }^{-2}(0) $ as a function 
of $T/T_{c }. $ Full line : theory from Eq.(9) with parameters given in 
the text. Filled circles : experimental results from Ref.[17].
b) Same as a) for $\lambda _{b }^{-2}(T) $ /
$\lambda _{b }^{-2}(0) $ as a function of $T/T_{c }$.}
\label{Fig1}
\end{figure}
\begin{figure}
\caption{$\lambda _{c }^{-2}(T) $ / $\lambda _{c }^{-2}(0) $ as a function
of    $T/T_{c }. $ Full line : theory from Eq.(9) with the same parameters
as in Fig.1. Filled circles : experimental results from
Ref.[32].}
\label{Fig2}
\end{figure}
\begin{figure}
\caption{a) $\lambda _{a }^{-2}(T) $ / $\lambda _{a }^{-2}(0) $
as a function   
of $T/T_{c }. $ Full line : theory from Eq.(9) for small plane-plane  
hopping (see text), with parameters given in the text. Filled  
circles : experimental results from Ref.[17].
b) Same as a) for $\lambda _{b }^{-2}(T) $ / 
$\lambda _{b }^{-2}(0) $   
as a function of $T/T_{c }. $ 
c) Same as a) for $\lambda _{c }^{-2}(T) $ /
$\lambda _{c }^{-2}(0) $   
as a function of $T/T_{c }. $ 
Filled circles : experimental results from
Ref.[32].}
\label{Fig3}
\end{figure}
 
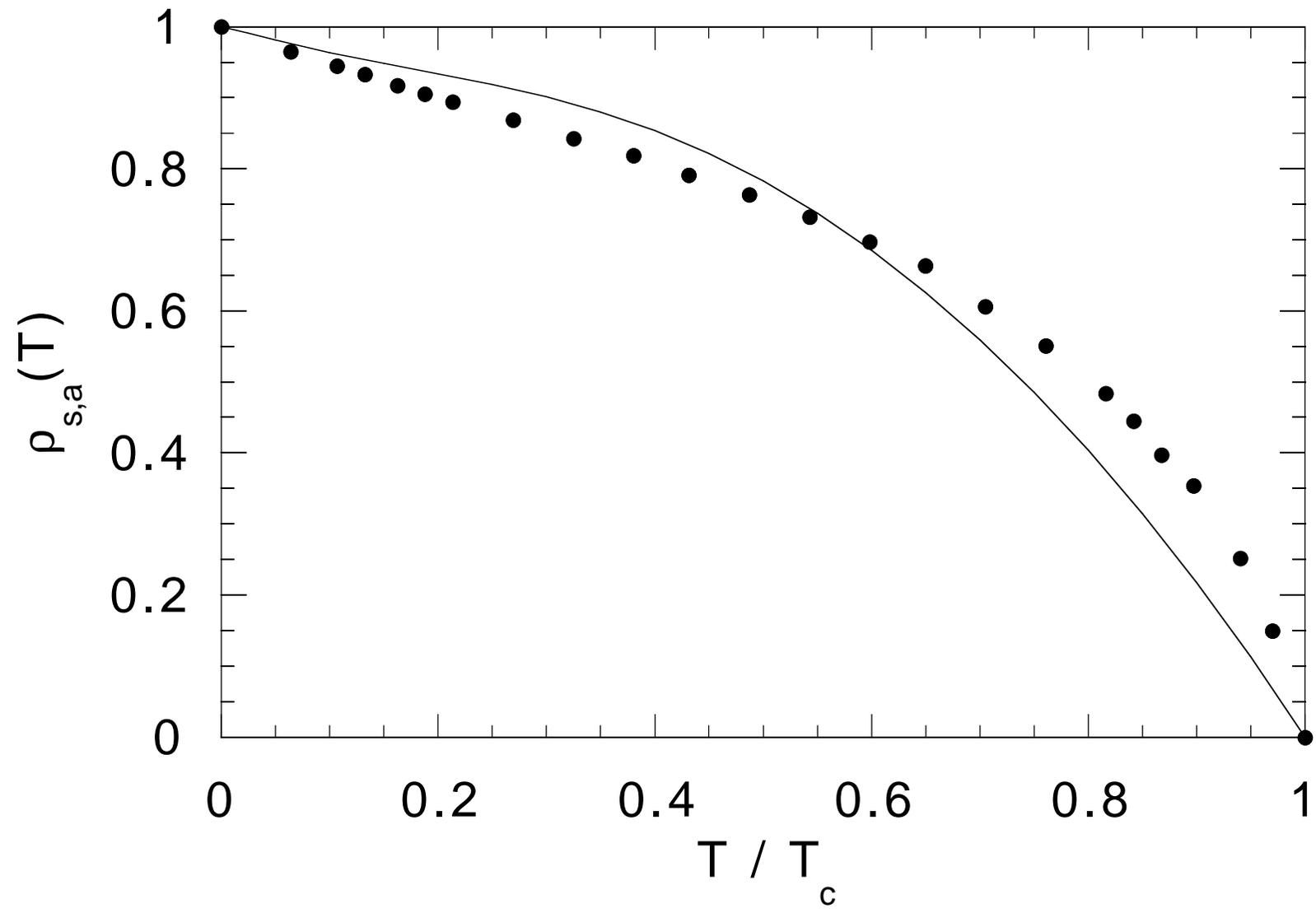
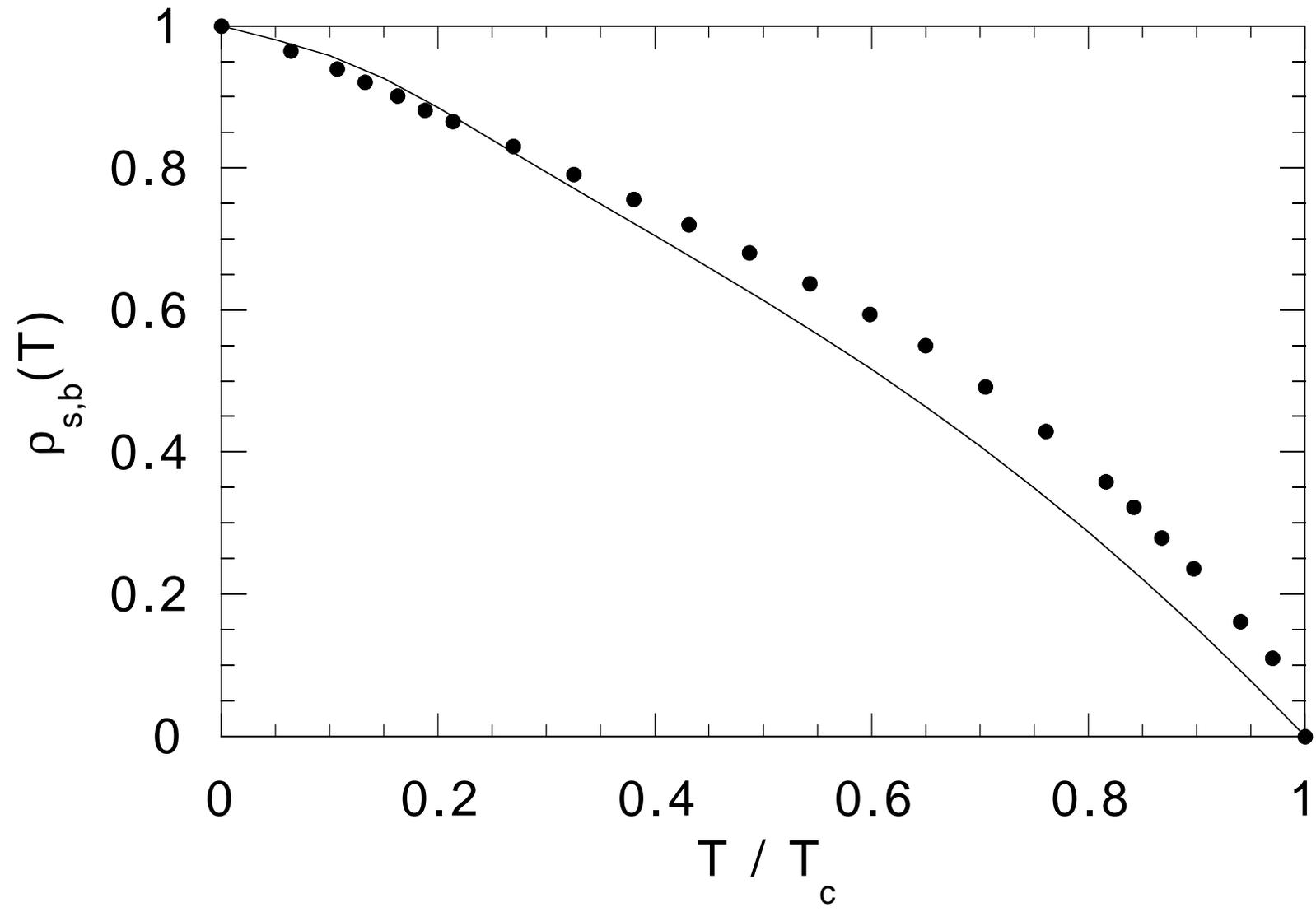
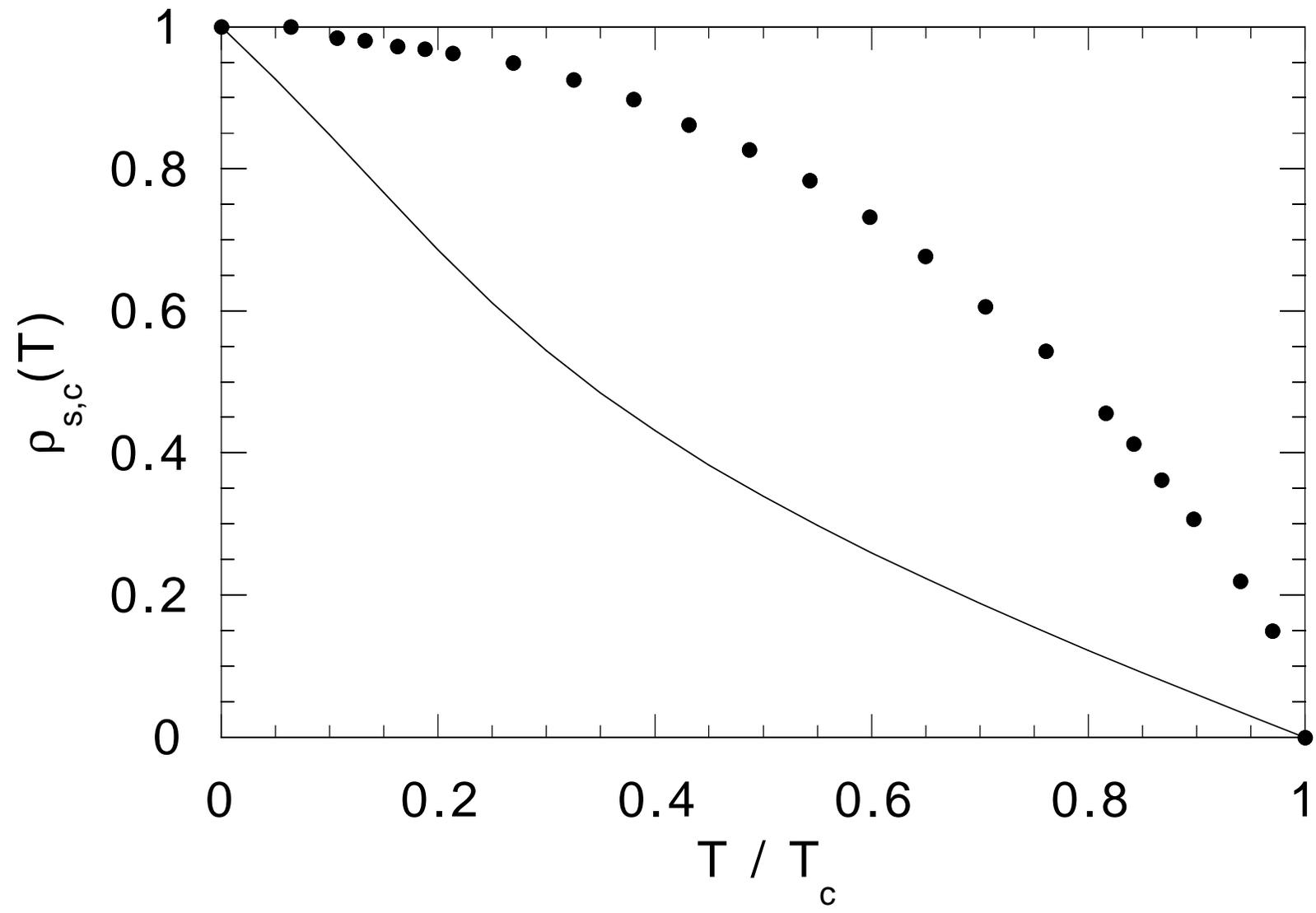
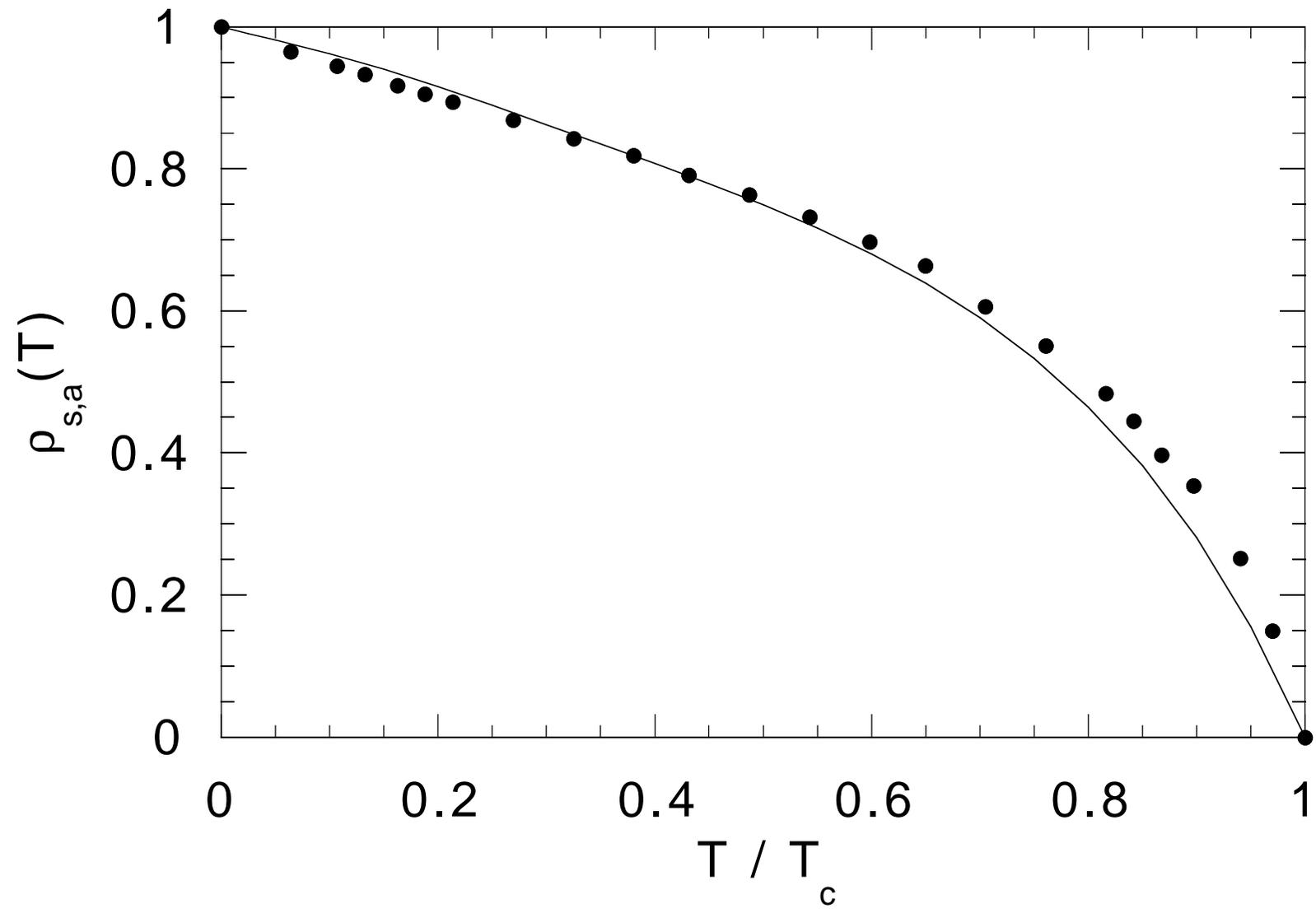
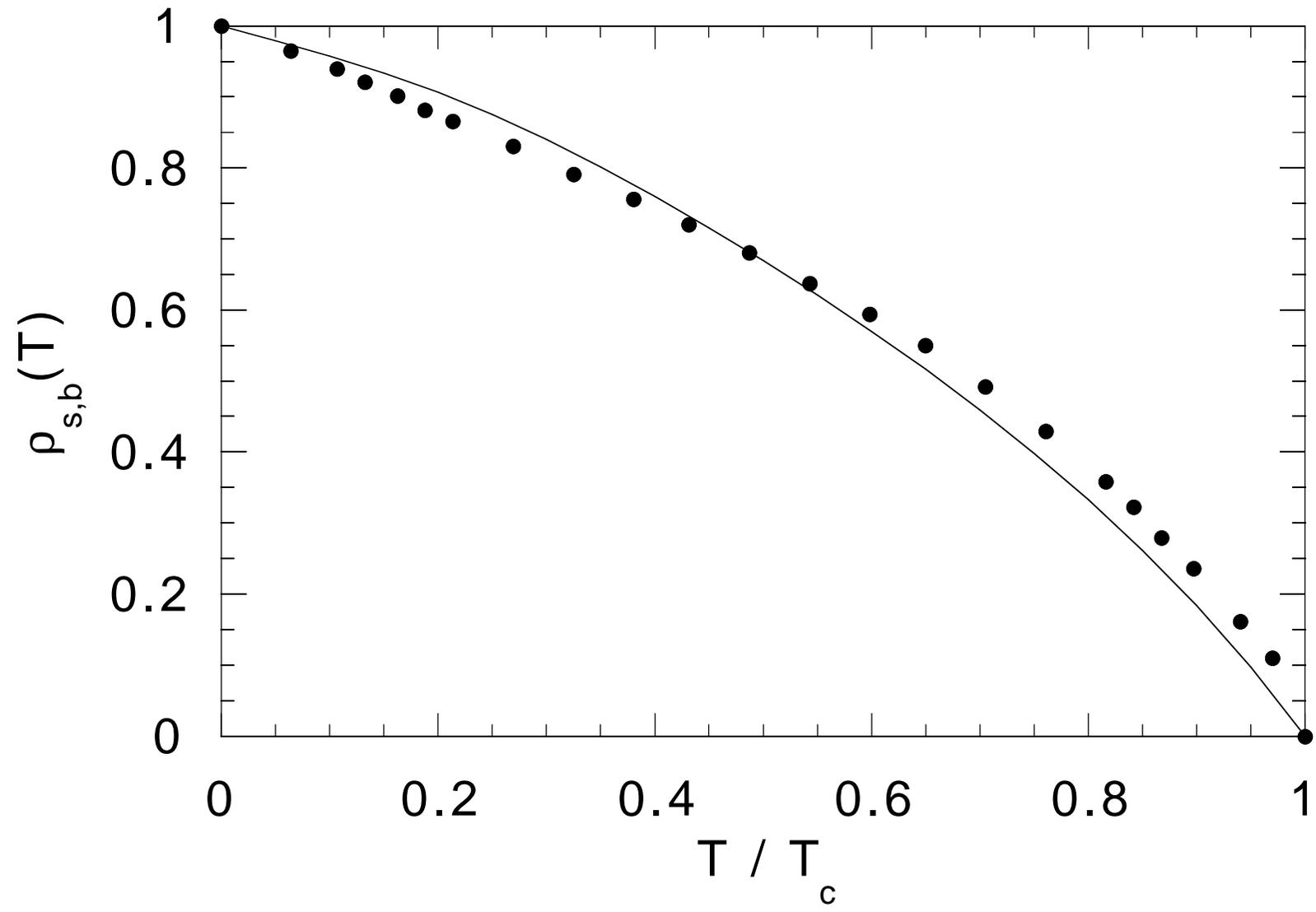
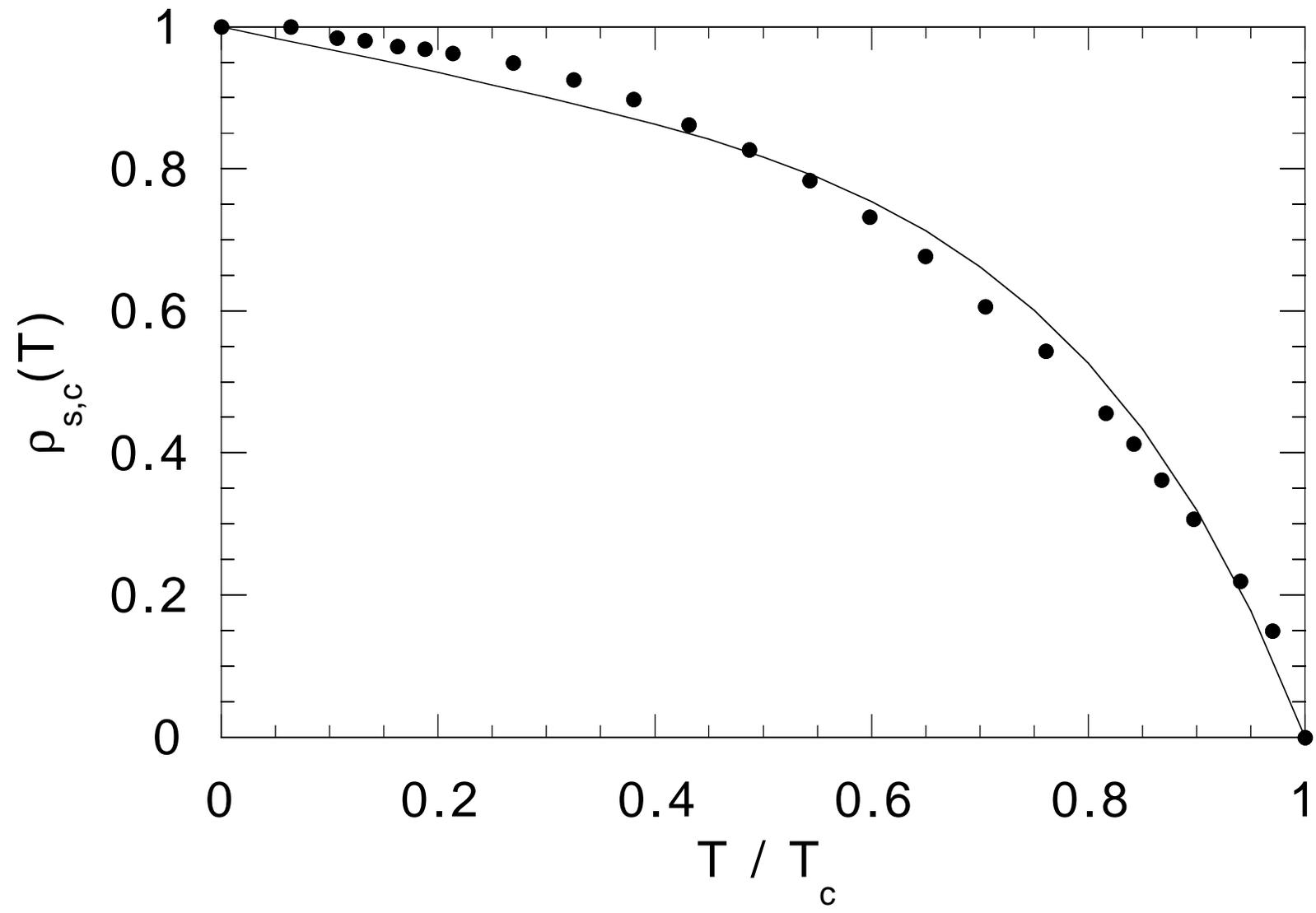
\end{document}